\documentclass[12pt]{iopart}

\usepackage{graphicx}

\begin{document}

\title[Growth kinetics of circular liquid domains]{Growth kinetics of circular liquid domains on vesicles by diffusion-controlled coalescence}

\author{Kazuhiko Seki$^1$, Shigeyuki Komura$^2$ and Sanoop Ramachandran$^3$}

\address{$^1$ National Institute of Advanced Industrial Science and Technology (AIST)\\
AIST Tsukuba Central 5, Higashi 1-1-1, Tsukuba, Ibaraki 305-8565, Japan}
\address{$^2$ Department of Chemistry, 
Graduate School of Science and Engineering, 
Tokyo Metropolitan University, 
Tokyo 192-0397, Japan}
\address{$^3$ Physique des Polym\`{e}res, Universit\'e Libre de Bruxelles,
Campus Plaine, CP 223, 1050 Brussels, Belgium
}

\begin{abstract}
Motivated by recent experiments on multi-component membranes, the growth kinetics 
of domains on vesicles is theoretically studied. 
It is known that the steady-state rate of coalescence cannot be obtained by 
taking the long-time limit of the coalescence rate when the membrane is regarded 
as an infinite two-dimensional (2D) system. 
The steady-state rate of coalescence is obtained by explicitly taking into 
account the spherical vesicle shape. 
Using the expression of the 2D diffusion coefficient 
obtained in the limit of small domain size, an 
analytical expression for the domain growth kinetics is obtained when the circular shape is always maintained. 
For large domains, 
the growth kinetics is discussed by investigating 
the size dependence of the coalescence rate 
using the expression for the diffusion coefficient of arbitrary domain size. 

\end{abstract}

\maketitle

 \section{Introduction}
 \label{sec:introduction}

Lipid bilayer membranes can be regarded as two-dimensional (2D) systems 
embedded in three-dimensional (3D) solvent. 
The membranes are coupled to solvent since the lipids composing the membrane interact with solvent surrounding it and the momentum can be exchanged between the membrane and the solvent~\cite{saffman-75,saffman-76,DeKoker,Komura2012,Fujitani,Seki,Ramachandran-10,Camley,Fan,Petrov}.
In this sense, membranes can be regarded as quasi-2D systems. 
By the recent advances in experiments, domains formed by phase separation 
in multicomponent membranes are visualized, and the lipid spatial organization 
and its dynamics have been 
studied~\cite{veatch-03,Baumgart-03,saeki-06,yanagisawa-07,Sezgin}.
The phase separation kinetics and coarsening are influenced by the dimensionality, 
domain shapes and the hydrodynamics in the systems~\cite{Bray}.
The phase separation in multicomponent lipid bilayer exhibits rich dynamics partly 
due to the momentum dissipation to the third-dimensions 
while the motion is confined to 2D~\cite{Ramachandran-10,Camley,Fan}.

In recent experiments, the growth kinetics of circular domains on ternary 
fluid vesicles has been observed by fluorescence 
microscopy~\cite{veatch-03,saeki-06,yanagisawa-07}.
In these experiments, liquid domains are formed in giant vesicles by phase 
separation into the 
liquid-ordered 
phase and the 
liquid-disordered 
phase on lowering the temperature from the one-phase region. 
Yanagisawa {\it et al.} found two different types of domain coalescence 
kinetics~\cite{yanagisawa-07}. 
In one of the coalescence processes, the domains grew by collision and coalescing 
while keeping the circular shape until the large domains of the vesicle size appeared. 
This growth kinetics due to the diffusion-controlled coalescence (DCC) was described 
by a power-law. 
In the other coalescence process, the domain growth was suppressed by membrane-mediated repulsive inter-domain interactions.
Recently, it was pointed out that the domain coalescence could be prevented 
by the membrane-mediated interactions between liquid domains associated with the 
deformations of the 
membrane~\cite{Lipowsky1992,Lipowsky2003,Seifert1997,Ursell2009,Semrau2009}.
When the liquid domain size exceeds a critical value, 
the boundary line energy is reduced by budding at the expense of some bending energy. 
The significant slowing down of the domain growth can be observed for large budded 
domains~\cite{yanagisawa-07,Semrau2009}. 
In the steady state, the domain patterns and the membrane shapes 
can be stabilized by the coupling between the local membrane curvature and the line 
tension~\cite{Baumgart-03,Sezgin,Andelman,Parthasarathy,Garcia-Saez}.

As briefly summarized above, the liquid domains coarsen under the influence 
of several competing processes.  
Even without budding, the observed growth kinetics was different from that 
obtained from the scaling hypothesis~\cite{saeki-06,yanagisawa-07}.
Motivated by the experiments, we study theoretically the growth kinetics 
of domains on vesicles by DCC.

The study of domain growth on vesicle surfaces is still limited compared to 
that in 3D~\cite{Bray}.
According to the scaling hypothesis, the domain growth exponent due to DCC  
in 2D is $1/2$ in contrast to that in 3D given by $1/3$~\cite{binder-74,Tomita}.  
However, it should be noted that, in the scaling hypothesis, the coalescence 
process is not explicitly taken into account. 
Moreover, it is known that the steady-state rate of coalescence cannot be 
obtained for an infinitely large 2D system~\cite{Naqvi,Barzykin}. 
In contrast to diffusion in 3D space, the boundary conditions 
are crucial to obtain the coalescence rate when the diffusion is restricted in 2D. 
This point is missing in the scaling hypothesis.

In addition to the above argument on pure 2D systems, the coalescence of 
liquid domains can be influenced by the coupling between the membrane and 
the solvent. 
The diffusion of large domains is more influenced by the coupling compared 
to that of small domains. 
As the domain grows, the influence of the coupling between the membrane and 
the solvent increases. 
The domain growth kinetics has been studied by dissipative particle dynamics 
simulations and continuum simulations~\cite{Ramachandran-10,Camley,Fan,LK04}.
The simulation results suggest the slowing down of the domain growth by DCC 
due to the coupling~\cite{Ramachandran-10}.

In this paper, we study the growth of liquid domains immersed in a 2D 
membrane by using an analytical theory which goes beyond the scaling hypothesis. 
We note that the steady-state rate of coalescence can be obtained by taking 
into account explicitly the vesicle shape~\cite{Sano,Bloomfield}.
By using the diffusion coefficient of domains obtained by taking into 
account the coupling between the membrane and the 
solvent~\cite{saffman-75,saffman-76,DeKoker,Komura2012,Fujitani,Seki,Petrov},
we show that it is independent of the domain size in the limit of small domain size. 
In such a case, it is known that the size distribution is described by the 
Smoluchowski theory of coalescence processes~\cite{Smoluchowski}. 
By further assuming that the circular shape of the liquid domains is always 
maintained, the time evolution equation of the mean domain size is obtained 
from the size distribution using the conservation of domain area upon coalescence. 
The mean domain growth is expressed by a single function for the whole time 
regime starting from the initial induction period of coalescence to the final 
asymptotic regime given by the power-law. 
When the domain size is large, we discuss the influence of the coupling 
between the membrane and the solvent on the domain growth by analyzing the 
size dependence of the coalescence rate.

In Sec.~\ref{sec:scaling}, we present the known results obtained from the 
scaling hypothesis.  
In Sec.~\ref{sec:Sinfinite}, the results of Smoluchowski theory in pure 
3D and 2D infinite systems are reviewed. 
In Sec.~\ref{sec:Svesicle}, the coalescence rate is obtained by taking 
into account the vesicle shape. 
In Sec.~\ref{sec:main}, the analytical expression representing the growth 
of mean domain size is obtained when the circular shape and the area of 
the liquid domains are kept before and after the coalescence. 
The size dependence of the the coalescence rate is investigated by using 
the analytical expression of the diffusion coefficient for the liquid 
domain of the arbitrary size in Sec.~\ref{sec:k2}.  
In Sec.~\ref{sec:discussion}, the theoretical results are discussed in 
relation to those obtained by the recent experiments~\cite{yanagisawa-07}.

\begin{figure}
\centerline{
\includegraphics[width=0.5\columnwidth]{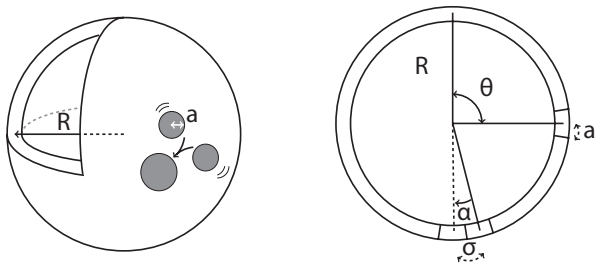}}
\caption{Schematic picture liquid domains embedded in a fluid vesicle (left). 
The radius of liquid domains is $a$ while the vesicle radius is $R$. 
Liquid domains undergo Brownian motion on the vesicle surface.
Shown on the right is the geometrical parametrization. 
$\theta$ denotes the azimuthal angle and 
$\alpha$ is the angle between domain centers at collision. 
$\sigma$ represents the encounter distance. } 
\label{fig:domain}
\end{figure}

\section{Scaling theory}
\label{sec:scaling}

In this section, we summarize the results obtained from the scaling theory. 
The scaling theory is the simplest way to derive the power-law growth of domain size. 
Obviously, one cannot obtain both the transient growth leading to the asymptotic 
power-law kinetics and the magnitude of the power-law growth.

The scaling theory is based on the hypothesis that 
the mean domain radius $\langle a(t) \rangle$ in $d$-dimension is related to 
time $t$ by 
\begin{equation}
\langle a(t) \rangle^2 \sim D_d t, 
\label{eq:scaling}
\end{equation}
where $D_d$ is the diffusion coefficient and subscript $d$ denotes the dimensionality. 
In 3D, the diffusion coefficient of the domain is inversely proportional to 
$\langle a(t) \rangle$ 
by the Stokes-Einstein relation, 
$D_3 \sim 1/\langle a(t) \rangle$. 
By substituting this relation in the scaling relation, 
we obtain $\langle a(t) \rangle \sim t^{1/3}$~\cite{Bray,binder-74,Tomita}.
For 2D, 
we have $D_2 \sim k_\mathrm{B} T/\eta$ by the dimensional argument, 
where  $k_\mathrm{B}$ is the Boltzmann constant, $T$ the temperature, 
$\eta$ the 2D membrane viscosity surrounding the domain.  
By using the above expression of $D_2$, 
the scaling hypothesis leads to 
$\langle a(t) \rangle \sim t^{1/2}$~\cite{Bray,binder-74,Tomita}.
It should be pointed out that 
a logarithmically time-dependent diffusion coefficient was derived in pure 
2D by the hydrodynamic theory~\cite{saffman-76}.
When the membrane couples to the solvent, 
a constant diffusion coefficient was derived, and 
$D_2$ depends logarithmically on the domain size
in the weak coupling limit~\cite{saffman-75,saffman-76}.
The logarithmic size dependence and the numerical factors are ignored in the dimensional argument.

We also note that the time given by $\langle a (t) \rangle^2/D_d$ is not equal to 
the coalescence time of the domain with size $\langle a (t) \rangle$. 
The coalescence time should be given by 
the mean first time that domains coalesce by diffusion from random initial distribution. 
Below, we show that the latter time is different from the one given by $\langle a(t)\rangle^2/D_d$.

The scaling hypothesis merely relates 
a single length scale given by the mean domain size at time $t$ with the diffusion coefficient as shown in  
Eq.~(\ref{eq:scaling}) and it should be justified. 
In the subsequent sections, we point out that 
the results of the scaling hypothesis cannot be obtained for the domain growth by the diffusion-coalescence in 
2D infinite systems. 
Then, we show that 
the scaling hypothesis is consistent with 
the domain growth kinetics by diffusion-coalescence 
when the available diffusion 
area is finite in 2D. 
The whole kinetics including transient growth and the final 
power-law growth will be obtained by taking into account the 
diffusion-coalescent process 
and the vesicle shape explicitly.

 \section{Smoluchowski theory in infinite systems}
 \label{sec:Sinfinite}

In the Smoluchowski theory, 
domain motion is assumed to be independent of the others   
and coalescence between a pair of domains is considered. 
When both domains can move, 
it is difficult  to solve the problem analytically. 
We assume that one of the domains is fixed and the other diffuses with the mutual diffusion coefficient 
$D_d$ which is the sum of the 
diffusion coefficients of two spherical domains of equal size~\cite{Rice}.
Coalescence takes place immediately when two domains come in contact 
at the encounter distance which is the sum of the radii of the two domains.  
In the theoretical formulation, the spatial distribution of domains satisfies the diffusion equation  
and the spatial domain density should vanish at the encounter distance. 
The domain size increases immediately after the coalescence and the spatial 
distribution of the new domain size should be zero at the 
encounter distance for the increased domain size.

The density of mobile domain around the immobile domain satisfies the diffusion equation 
\begin{equation}
\frac{\partial}{\partial t} \rho_d (r,t) = D_d \nabla^2 \rho_d(r,t),
\label{eq:suv1}
\end{equation}
where $r$ is the distance to the center of the immobile domain. 
We assume random initial condition given by, 
\begin{equation}
\rho_d (r,t=0)=1. 
\label{eq:suv2}
\end{equation}
The boundary condition applied at the encounter distance 
$\sigma$ is given by 
\begin{eqnarray}
\rho_d(r=\sigma,t)=0. 
\label{eq:suv3}
\end{eqnarray}
We should set another boundary 
condition such that the density at infinite separation is unity, i.e., 
\begin{equation}
\lim_{r \rightarrow \infty} \rho_d (r,t)=1. 
\label{eq:suv4}
\end{equation}

\subsection{Three-dimensions (3D)}
Before investigating the domain growth on the 2D spherical 
surface, we shall briefly present the known results for 3D 
infinite systems and show that the corresponding results do not 
hold for 2D cases. 

For 3D infinite systems, 
the density profile which satisfies both of the boundary conditions is 
$\rho_3 (r)= 1 - (\sigma/r)$~\cite{Rice}.
The mean coalescence rate is given by the inward flow of domains 
across the surface at $\sigma$
\begin{equation}
k_3 = 4 \pi \sigma^2 D_3
\left( \frac{\mathrm{d} \rho_3(r)}{\mathrm{d} r} \right)_{r=\sigma}
=4 \pi \sigma D_3. 
\label{eq:suv5}
\end{equation}
In a 3D fluid of viscosity $\eta_{\rm s}$, 
the diffusion coefficient is inversely proportional to the size of the diffusing object, 
$D_3 = k_\mathrm{B} T/3 \pi \eta_\mathrm{s} \sigma$. 
By using this Stokes-Einstein relation,  
the mean coalescence rate can be expressed  
by $k_3=4 k_\mathrm{B} T/3 \eta_\mathrm{s}$.
Notice that $\sigma/2$  is the radius of the spherical domain~\cite{Smoluchowski}.

 \subsection{Two-dimensions (2D)}
 
By applying the boundary condition at $\sigma$, we obtain 
\begin{equation}
\rho_2(r)=C \ln(\sigma/r), 
\label{eq:rho2}
\end{equation}
where $C$ is a constant determined 
from the other boundary condition at $r \rightarrow \infty$. 
However, it is impossible to fix $C$ because $\rho_2(r)$ diverges 
for $r \rightarrow \infty$ as long as $C$ is finite. 
Unlike 3D infinite systems, 
the density cannot satisfy both boundary conditions, 
Eqs.~(\ref{eq:suv3}) and (\ref{eq:suv4}), simultaneously.

 \section{Smoluchowski theory in spherical surface}
 \label{sec:Svesicle}

The difficulty mentioned in the previous section for 2D can be overcome if the 
available diffusion area is finite.
The density profile depends crucially on the shape of the 2D region. 
In the experiments~\cite{yanagisawa-07}, DCC 
of circular domains was observed on the vesicle surfaces. 
In principle, 
the coalescence rate can be obtained by modifying the method shown in the previous section 
applicable to the spherical region 
but the calculation is rather complicated. 
In this paper, we employ an alternative method. 

In the method, 
we investigate the life time of 
the density of mobile domains survived from collision 
to the immobile domain.  
In a confined region, the decay of the density can be well approximated by a single exponential.  
The time constant of the exponential decay can be reasonably obtained from the mean first-passage time of a mobile domain to the periphery of the immobile domain  
by assuming uniform distribution for the starting point. 
For the coalescence, the mean first-passage time is the mean coalescence time 
corresponding to the encounter time between two domains. 
When the initial position of the mobile reactant is $z=\cos \theta$ 
(see Fig.~\ref{fig:domain} for the geometry), 
the mean coalescence time $\tau(z)$ satisfies the following equation
(see the Appendix for the derivation)~\cite{Sano}
\begin{equation}
\frac{D_2}{R^2} \frac{\partial}{\partial z} (1-z^2) \frac{\partial}{\partial z} \tau (z)= -1. 
\label{eq:mft1}
\end{equation}
The boundary conditions are  
\begin{equation}
\tau (z=-\cos \alpha)=0,~~~~~  
\left( \frac{\partial \tau (z)}{\partial z} \right)_{z=1} =0, 
\label{eq:mft2}
\end{equation}
where $\alpha$ is the angle between domain centers at collision (see Fig.~\ref{fig:domain}).  
Notice that a simple geometric argument gives 
$\sin (\alpha/2)= \sigma/(2R)$ and  
$\cos \alpha= 1 - \sigma^2/(2R^2)$~\cite{Sano}.
Equations~(\ref{eq:mft1}) and (\ref{eq:mft2}) can be easily solved, and we 
obtain 
\begin{equation}
\tau (z) =\frac{R^2}{D_2} \ln \left[ \frac{2 R^2}{\sigma^2} (1+z) \right]. 
\label{eq:tauz}
\end{equation}
By averaging over the random initial distribution, we obtain~\cite{Sano}
\begin{equation}
\tau_\mathrm{av} = \frac{\int_{-\cos \alpha}^1 {\rm d}z \, \tau (z) }{2- \sigma^2/(2R^2)}
=\frac{R^2}{D_2} \left[ \frac{2}{1-(\sigma/2R)^2} \ln \left(
\frac{2R}{\sigma} \right) -1 \right]. 
\label{eq:tau}
\end{equation}
The probability that the mobile domain has not reached the immobile domain up to time $t$ is given by  
$\exp \left( -t/\tau_\mathrm{av} \right)$ when 
domains are initially distributed uniformly in the spherical surface. 
The same probability can be expressed by using the bulk bimolecular rate as
$\exp \left( -k_d t/A \right)$, 
where the surface area available for diffusion is 
$A=2 \pi R^2 [2- \sigma^2/(2R^2)]$. 
Comparing these expressions, we find that 
the bulk bimolecular rate can be calculated from the mean coalescence time by 
$k_2=A/\tau_\mathrm{av}$~\cite{Sano,Bloomfield,Tachiya83}.

With the use of the bimolecular rate, the mean coalescence rate 
is obtained as~\cite{Sano,Bloomfield,Tachiya83}
\begin{eqnarray}
k_2&=\frac{4 \pi [1-(a/R)^2 ] D_2(a)}{2 \ln (R/a) -1+(a/R)^2}
\label{mft2dim0}
\\
&\simeq \frac{4 \pi D_2(a)}{2 \ln (R/a) -1}, 
\label{mft2dim}
\end{eqnarray}
where $a \ll R$ is used to obtain the second equality.  
The mutual diffusion coefficient is now expressed 
by $D_2(a)$ since it depends on
the domain radius $a$ as shown below.

Unfortunately, the full size-dependence of the diffusion 
coefficient is not known for spherical vesicles.
However, the analytical expression is known for a circular liquid domain which 
has the same viscosity as the outside of the domain for 2D flat membranes. 
In this case, the mutual diffusion coefficient was obtained by 
De Koker as~\cite{DeKoker}
\begin{equation}
D_2(a) = \frac{2 k_\mathrm{B} T}{\pi \eta } \int_0^\infty \mathrm{d} z \,
\frac{J_1^2(z)}{z^2 (z + \nu a )},   
\label{eq:D}
\end{equation}
when the two circular domains have the same radius $a$. 
In the above, 
$J_1(z)$ is the Bessel function of the first kind, 
$\eta$ is the 2D membrane viscosity, 
$\nu= 2 \eta_\mathrm{s}/\eta$ with $\eta_\mathrm{s}$ being the viscosity of the 
outer fluids. 
Here we have assumed that the viscosities of the liquid inside and outside 
the vesicle are the same.
The analytical expression after 
the integration can be expressed using Meijer 
$G$-functions~\cite{Seki,Fujitani}. 
In the case of $\nu a \ll 1$, the above expression reduces to~\cite{Seki,Fujitani}
\begin{equation}
D_2(a) \approx \frac{k_\mathrm{B} T}{\pi \eta}
\left[\ln\left(\frac{2}{\nu a}\right)-\gamma +\frac{1}{4} \right],  
\label{eq:dES}
\end{equation} 
where
 $\gamma=0.5772 \cdots$ is Euler's constant. 
Equation (\ref{eq:dES}) is slightly larger than the mutual diffusion coefficient 
of the Saffman-Delbr\"uck (SD) theory derived for solid domains under the condition of $\nu a \ll 1$. 
Strictly speaking, the SD result was also obtained for 2D flat membranes. 
Recently, it was shown that the SD result was applicable for spherical vesicles
when $a < R < 1/\nu$~\cite{Henle}.

By substituting  Eq.~(\ref{eq:dES}) into Eq.~(\ref{mft2dim}), we obtain 
\begin{equation}
k_2 \approx \frac{k_\mathrm{B} T}{\eta}, 
\label{Rmean_coagulation_rate1}
\end{equation}
when $a/R \ll1$. 
The coalescence  rate is independent of $a$ in this limit. 
Then $k_2 t$ approximately represents the area explored by a diffusive object of 
radius $\sigma$ during time $t$~\cite{Tachiya83,Oshanin}. 
Hence $k_2 t$ is given by $D_2(a) t$ times the effective collision cross-section. 
The size independence is the result of the two opposing effects;   
with increasing the domain size, the diffusion coefficient decreases while 
the effective collision cross-section given by $4 \pi/[2 \ln(R/a)-1]$ in 
Eq.~(\ref{mft2dim}) increases. 

Strictly speaking, the coalescence rate depends on the domain size if the 
coalescence occurs between domains having different sizes.  
However, such a size dependence is small for 2D because of the weak 
logarithmic dependence in Eqs.~(\ref{mft2dim}) and (\ref{eq:dES}), and 
will be ignored hereafter.
In fact, the size dependence was not taken into account even to study the 
coalescence processes in 3D~\cite{Smoluchowski}.

Equation (\ref{mft2dim0}) is valid even if the condition $a/R \ll1$ is not
satisfied. 
The full size-dependence of the coalescence rate will be studied 
by substituting Eq.~(\ref{eq:D}) into Eq.~(\ref{mft2dim0}) as we shall
discuss in Sec.~\ref{sec:k2}.

 \section{Growth kinetics of liquid domains}
\label{sec:main}
 
Now we consider the formation of $m$-fold domains from the initial domains 
with the same size. 
When the coalescence rate is independent of the domain size as discussed 
above, the number density of $m$-fold domain $n_m(t)$ at time $t$ is given 
by~\cite{Smoluchowski} 
\begin{equation}
\frac{n_m(t)}{n_1} = 
\frac{\left(t/\tau\right)^{m-1}}{(1+t/\tau)^m},  
\label{k_aggregate}
\end{equation}
where 
\begin{equation}
\tau=\frac{1}{k_d n_1}, 
\label{tF}
\end{equation}
and $n_1$ is the initial number density of the primary domains, and 
$k_d$ the coalescence rate in $d$-dimensional space.

\begin{figure}%
\centerline{
\includegraphics[width=0.7\columnwidth]{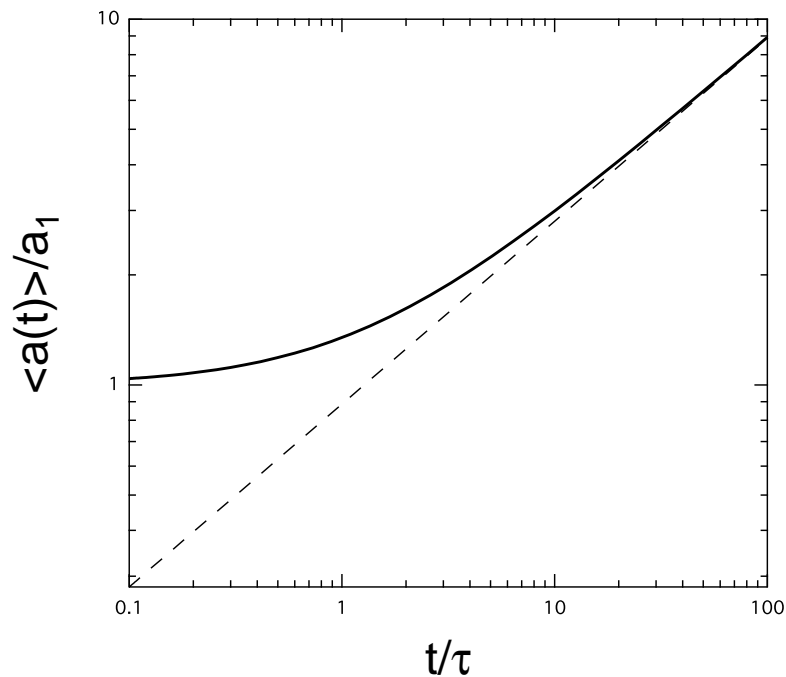}
}
\caption{The average domain size $\langle a(t) \rangle/a_1$ 
as a function of dimensionless time $t/\tau$ when domains are confined 
in the spherical surface. 
The solid line is obtained from Eq.~(\ref{aveRt}) and  
the dashed line represents the asymptotic behavior given by Eq.~(\ref{aveRt_asym}).  
} 
\label{fig:domaingrowth_r}
\end{figure}

In general, 
if the aggregates of large sizes are in the solid state, 
there are many possible shapes of aggregates. 
The shape characterization and 
the size distribution have been studied in the conventional theory of Brownian coagulation. 
Here, we study Brownian coalescence of liquid domains under the condition that 
the circular shape is always maintained. 
For comparison, 
we also consider Brownian coalescence of liquid domains in 3D infinite systems, 
where the spherical shape is always maintained. 
Under this assumption, 
the time evolution of the domain size can be obtained analytically. 

By assuming that the domain shape is immediately restored after the coalescence, 
the radius of the $m$-fold domain $a_m$ is determined by 
the area ($d=2$) or volume ($d=3$) conservation relation, 
\begin{eqnarray}
a_m^d = m a_1^d.
\label{radius}
\end{eqnarray}
By averaging $a_m$ over the distribution of $m$-fold aggregate [see 
Eq.~(\ref{k_aggregate})], the mean domain radius $\langle a(t) \rangle$ 
at time $t$ is obtained as  
\begin{equation}
\frac{\langle a(t) \rangle}{a_1} = 
\sum_{m=1}^\infty m^{1/d} \frac{n_m (t)}{n_1} = 
\frac{\tau}{t} \mbox{Li}_{-1/d} \left(
\frac{t/\tau}{1+t/\tau} 
\right).  
\label{aveRt}
\end{equation}
In the above, we have used the polylogarithm function defined 
by~\cite{abram-stegun}
\begin{equation}
\mbox{Li}_s (x) = \sum_{k=1}^\infty \frac{x^k}{k^s}. 
\end{equation}

In the asymptotic limit of $t/\tau \gg 1$, Eq.~(\ref{aveRt}) reduces to 
either 
\begin{equation}
\frac{\langle a(t) \rangle}{a_1} \approx
\frac{\sqrt{\pi}}{2} \left(\frac{t}{\tau} \right)^{1/2} 
\label{aveRt_asym}
\end{equation}
for the coalescence in 2D spherical surfaces, or  
\begin{equation}
\frac{\langle a(t) \rangle}{a_1} \approx
\Gamma \left(\frac{4}{3}\right) \left(\frac{t}{\tau} \right)^{1/3}
\label{aveRt_3d}
\end{equation}
for that in 3D infinite systems. 
Here $\Gamma(x)$ is the gamma function. 
The time evolution of 
$\langle a(t) \rangle/a_1$ is shown in Fig.~\ref{fig:domaingrowth_r} for 
the domain coalescence on the vesicle. 
The asymptotic time dependence is well approximated by Eq.~(\ref{aveRt_asym}) when 
$t/\tau \gg 1$. 
We can see the induction period of coalescence when $t/\tau < 1$. 
The induction period is characterized by the inverse of the apparent coalescence rate 
given by the coalescence rate times 
the initial number density of the primary domains, 
Eq. (\ref{tF}).  
By taking into account explicitly the vesicle shape, 
the finite coalescence rate is obtained from the Smoluchowski theory and the results show 
the induction period before the asymptotic growth.

 \section{Size dependence of the coalescence rate}
 \label{sec:k2}

\begin{figure}%
\centerline{
\includegraphics[width=0.7\columnwidth]{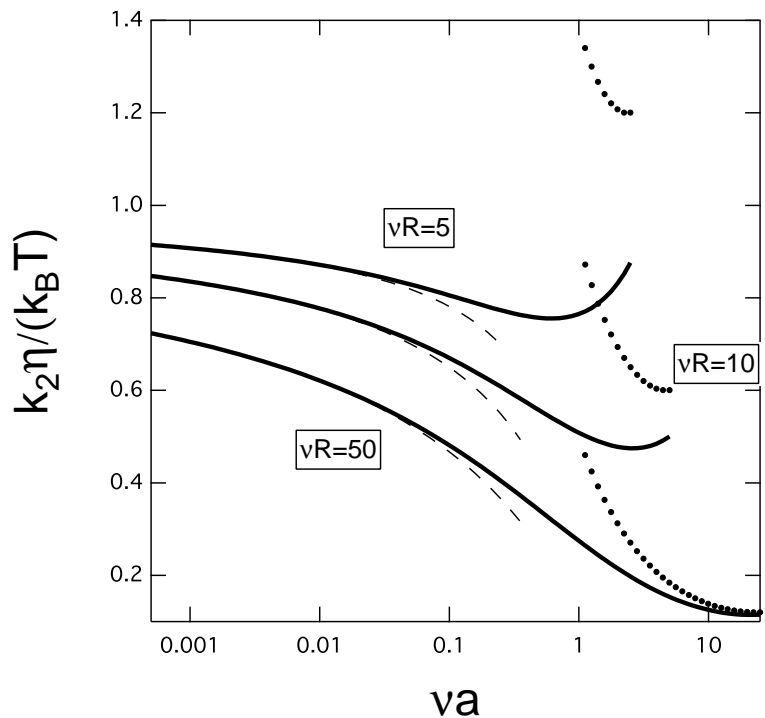}
}
\caption{
$k_2 \eta/\left(k_\mathrm{B} T\right)$ as a function of 
the dimensionless domain size $\nu a$. 
The solid line is obtained from Eq.~(\ref{mft2dim0}) using the expression of 
the diffusion coefficient for a circular liquid domain which has the same viscosity 
as the outside of the domain. 
$R \nu =5$, $10$ and $50$ from top to bottom. 
The dashed lines are obtained by substituting Eq.~(\ref{eq:dES}) into 
Eq.~(\ref{mft2dim0}).  
The dots are obtained by substituting Eq.~(\ref{eq:dLES}) into Eq.~(\ref{mft2dim0}). 
} 
\label{fig:k2_r}
\end{figure}

In the previous section, the growth kinetics of the circular liquid domain by diffusion 
coalescence is obtained when the coalescence rate is independent of the size of the 
liquid domain. 
The result for the domain growth on the vesicle is obtained in the limit of $\nu a \ll 1$. 
Below, we study the size dependence of the coalescence rate using 
Eqs.~(\ref{mft2dim0}) and (\ref{eq:D}). 
The diffusion coefficient Eq.~(\ref{eq:D}) is obtained for a circular liquid domain 
of arbitrary size. 
In the limit of $\nu a \ll 1$, 
the domain size dependence of the diffusion coefficient is logarithmic  
as shown by Eq.~(\ref{eq:dES}). 
The coalescence rate is given by $k_\mathrm{B}  T/\eta$ 
independent of the domain size because of the cancellation of the 
logarithmic domain size dependence both in 
Eqs.~(\ref{mft2dim0}) and (\ref{eq:dES}). 
In the opposite limit of $\nu a \gg 1$, 
the diffusion coefficient is obtained as~\cite{Fujitani,Seki}
\begin{equation}
D_2(a) \approx  \frac{8 k_\mathrm{B} T}{3\pi^2 \eta_\mathrm{s} a}, 
\label{eq:dLES}
\end{equation}
which is inversely proportional to the domain radius $a$.

The coalescence rate obtained by substituting Eq.~(\ref{eq:D}) into 
Eq.~(\ref{mft2dim0}) is shown by the solid lines in Fig.~\ref{fig:k2_r}. 
For relatively small vesicle, say $\nu R =5$, the coalescence rate is almost 
independent of the domain size and is approximately given by $k_\mathrm{B}  T/\eta$. 
The situation is very different for large vesicles. 
When $\nu R \geq10$, the size dependence of the coalescence rate can be ignored 
only when $\nu a <0.1$. 
In the same figure, we show the results obtained by substituting Eq.~(\ref{eq:dES}) 
into Eq.~(\ref{mft2dim0}) using dashed lines. 
As long as the the domain size dependence of the diffusion coefficient can be 
approximated by Eq.~(\ref{eq:dES}), 
the size dependence of the coalescence rate is weak.

When $\nu R =5$, the size dependence is weak 
even beyond the regime given by $\nu a \ll 1$. 
However, when $\nu R \geq 10$ and $\nu a>0.1$  
the coalescence rate decreases by increasing the domain size. 
In the figure, we show the results obtained by substituting 
Eq.~(\ref{eq:dLES}) into Eq.~(\ref{mft2dim0}) using dots. 
The coalescence rates with the full size-dependence of the diffusion 
coefficients [Eq.~(\ref{eq:D})] approach the results shown by dots as 
the size increases. 
The dots rapidly decrease by increasing the domain size because of the power-law 
dependence of the domain size in Eq.~(\ref{eq:dLES}). 
When $\nu R \geq 10$ and $\nu a > 0.1$, the coalescence rate decreases by increasing 
the domain size because of the strong size dependence of the diffusion coefficient. 
If the coalescence rate decreases by increasing the size, 
the growth kinetics of the domain size should be suppressed. 
The growth law given by Eq.~(\ref{aveRt}) and the result shown in 
Fig.~\ref{fig:domaingrowth_r} can be applicable when $\nu R =5$, but should be 
applicable only when $\nu a \ll1$ if the vesicle radius is large such as 
$\nu R \geq 10$.

We also note that the coalescence rate increases by increasing the domain size 
if the domain size becomes close to the vesicle radius. 
This dependence originates from the denominator of Eq.~(\ref{mft2dim0})  
and can be interpreted as the finite size effect of the domains confined 
in the spherical surface.

The diffusion coefficient Eq.~(\ref{eq:D}) shows the logarithmic 
size dependence [Eq.~(\ref{eq:dES})] and the power-law size dependence 
[Eq.~(\ref{eq:dLES})] as $\nu a$ is varied. 
The dimensionless quantity $\nu a$ characterizes the coupling 
between the embedding bulk fluid and the membrane for given size. 
The coupling is small when $\nu a \ll1$. 
The domain growth and its asymptotic limit given by Eqs.~(\ref{aveRt})
and (\ref{aveRt_asym}), respectively, are obtained in the weak coupling limit. 
In the strong coupling limit $\nu a \gg1$, the domain growth is suppressed 
as a result of the small coalescence rate compared to that in the weak coupling 
limit as shown in Fig. \ref{fig:domaingrowth_r}. 
The results are consistent with the simulation results that the growth law is 
suppressed by increasing the hydrodynamic coupling when the governing mechanism 
is DCC in 2D~\cite{Ramachandran-10,Camley}.

The above conclusion is not altered if we use the diffusion coefficient for 
the solid domains instead of Eq.~(\ref{eq:D}). 
Recently, the diffusion coefficient of the solid domain in the 2D flat membrane 
is approximated by a closed-form empirical expression~\cite{Petrov}.
The interpolation formula of the diffusion coefficient reproduces the logarithmic 
size dependence obtained from SD theory when $\nu a \ll 1$. 
When the interpolation formula is introduced, 
the coalescence rate Eq.~(\ref{mft2dim0}) is slightly smaller but the overall 
size dependence is similar to that obtained by using Eq.~(\ref{eq:D}).
The coalescence rate decreases by the decrease of the diffusion coefficient of the 
solid domain compared to that of the liquid domain of the same size. 
In general, the diffusion coefficient of the solid domain is smaller than that of the liquid domain of the same size 
since the friction between the membrane and the solid edge is larger than that between the 
membrane and the liquid domain.

As shown above, the coalescence rate is constant in time when domains are confined 
on a spherical vesicle. 
The finite coalescence rate constant can also be calculated by considering a 
circular flat sheet of radius $L$ by setting the absorbing boundary condition 
at $r=a$ and the reflecting boundary condition at $r=L$ ($a<L$).
When a circular domain of radius $a$ is placed at the center of a circular 
flat sheet, the coalescence rate is given by Eq.~(\ref{mft2dim}) by replacing 
$R$ with $L$~\cite{Adam}. 
Within this substitution, the domain size dependence of the coalescence rate 
is the same as that in a spherical vesicle. 
The diffusion coefficient of a domain placed at the center of a circular sheet 
of radius $L$ was obtained by taking into account the viscosity of a circular 
flat membrane~\cite{saffman-75}. 
However, the result was limited for $a \ll L$ and obtained by ignoring the 
hydrodynamic coupling between the membrane and the solvent.

 \section{Discussion and conclusion}
 \label{sec:discussion}
 
We discuss the relevant length scales in the recent experiment on domain growth kinetics by diffusion coalescence~\cite{yanagisawa-07}.
In the experiments, 
giant vesicles with a diameter of about $20$ $\mu$m 
undergo phase separation  
at 30 $^\circ$C after the temperature 
drop from the one-phase region (42--43 $^\circ$C). 
The circular domains were observed when the size exceeded an optical resolution of the microscope (roughly $0.8$ $\mu$m). 
In one of the coalescence processes, 
the large domains of the vesicle size appeared within several minutes. 
The domain growth by collision and coalescence was observed. 
In the other coalescence process, the domain growth was suppressed for several $10$ minutes.
The former process can be theoretically studied by assuming DCC.

The value of $\nu=2 \eta_\mathrm{s}/\eta$ can be estimated by using the typical values of 
$\eta_\mathrm{s}=10^{-3}$ Pa$\cdot$s, and $\eta$ given by $0.1$ Pa$\cdot$s times the membrane thickness 
$5$ nm as $\nu=4.0 \times 10^{6}$ $\rm{m}^{-1}$. 
We estimate $\nu R \approx 40$ by introducing 
the typical radius of the vesicles $10$ $\mu$m. 
When $\nu R =50$, the coalescence rate is almost independent of the domain size 
as long as $a< 25$ nm obtained from the condition of $\nu a < 0.1$. 
The domain size is much smaller than the optical resolution of the microscope such as
$0.8$ $\mu$m. 
Therefore, when the domain size grows and reaches to the optical resolution,   
the coalescence rate decreases with increasing the domain size and the domain growth 
can be suppressed compared to that given by Eq.~(\ref{aveRt_asym}).  
According to Fig.~\ref{fig:k2_r}, 
the coalescence rate is almost constant over the wide range of the domain size when $\nu R =5$. 
This value corresponds to the vesicle radius close to $1$ $\mu$m which 
may be the maximum vesicle radius to observe the power-law growth given by 
Eq.~(\ref{aveRt_asym}).

It should be remembered that the full size-dependence of the diffusion 
coefficient for spherical vesicles is not known. 
In this paper, the size dependence of the coalescence rate has been discussed 
by substituting the known expression of the diffusion coefficient for 2D flat 
membranes. 
The full size-dependence of the diffusion coefficient for the spherical vesicle 
is needed to further develop the coagulation theory for the large domains 
within the current limit of optical resolutions.

In the original work by Yanagisawa \textit{et al.}, the best fitted exponent 
2/3 was obtained for the power-law domain growth~\cite{yanagisawa-07}.
However, the mechanism which leads to this large exponent is not well-understood. 
An attractive interaction between domains seems to be present due to the 
hydrodynamic flow around domains, which would accelerate the domain growth~\cite{Taniguchi}.

Power law growth of circular domains can be induced by 
transport of molecules from one domain to another through the 
medium~\cite{Bray,Lifshitz,Miguel,Ehrig}.
The growth of large domains is associated with evaporation of small domains, 
which is known as Ostwald ripening. 
According to the Lifshitz-Slyozov-Wagner theory of Ostwald ripening,  
the power law exponent is $1/3$~\cite{Lifshitz}.
In this paper, we studied the domain growth by the DCC mechanism 
and did not consider the evaporation and condensation mechanism in 2D. 
We just remark that the exponent $1/3$ 
is independent of the dimensionality and holds also in 2D~\cite{Camley,Miguel,Ehrig}.

In conclusion, we have investigated DCC mechanism for growth 
kinetics of the liquid domains on the fluid vesicles.
By applying the bimolecular reaction theory in the spherical surface and using the 
the 2D diffusion coefficient, the 2D coalescence rate is found to be 
independent of the liquid domain size if it is small enough.
As a result, the domain size distribution is given by the classical 
Smoluchowski theory. 
When the circular shape is always maintained, 
we have obtained the mean domain size for 
the whole time range [Eq.~(\ref{aveRt})] by using the domain size distribution and 
the area conservation relation. 
In the asymptotic long time limit, we expect the power-law behavior with the 
exponent 1/2 [Eq.~(\ref{aveRt_asym})]. 

The domain growth kinetics has been derived under the condition that the 2D coalescence rate is independent of the domain size. 
The condition is investigated 
by using recently obtained analytical expression for the diffusion coefficient of arbitrary domain size. 
When the vesicle radius is small, the coalescence rate can be well-approximated as a constant over the wide range of 
the domain size. 
When the vesicle radius is large,  
the coalescence rate becomes independent of the domain size only in the limit of the small domain size. 
In general, the coalescence rate decreases by increasing the domain size up to a certain size where  
the finite size effect dominates. 
The results are discussed in relation to the recent experimental observations of DCC in vesicles.

\ack

We would like to thank M. Imai and M. Yanagisawa for valuable discussions. 
KS and SK are supported by Grant-in-Aid for Scientific Research
(grant No.\ 24540439) from the MEXT of Japan.
SK also acknowledges the supported by the JSPS Core-to-Core Program
``International research network for non-equilibrium dynamics 
of soft matter''.

\appendix
\section{The mean coalescence time}

In this Appendix, we briefly present the derivation of Eq.~(\ref{eq:mft1})  
when the density satisfies Eqs.~(\ref{eq:suv1})--(\ref{eq:suv4}). 
The density can be expressed using the probability of finding a pair of domains at 
the relative position $\bi{r}$ at time $t$ 
if their initial relative position was $\bi{r}_{\rm i}$ and was uniformly distributed 
\begin{equation}
\rho_d ( \bi{r}, t) = \int {\rm d} \bi{r}_{\rm i} \, 
p_d(\bi{r}, t | \bi{r}_{\rm i}, 0). 
\label{eq:appendix1}
\end{equation}
We introduce the survival probability  
that the pair has not coalesced up to time $t$ if 
their initial relative position was $\bi{r}$
 \begin{equation}
w_d ( \bi{r}, t) = \int {\rm d} \bi{r}_{\rm f} \, 
p_d(\bi{r}_{\rm f}, t | \bi{r}, 0). 
\label{eq:appendix2}
\end{equation}
Because $p_d(\bi{r}_{\rm f}, 0 | \bi{r}, -t)$ satisfies the backward Kolmogorov 
equation and hence $p_d(\bi{r}_{\rm f}, t | \bi{r}, 0)=
p_d(\bi{r}_{\rm f}, 0 | \bi{r}, -t)$, 
we obtain~\cite{Weiss,Risken,Gardiner} 
\begin{equation}
\frac{\partial}{\partial t} w_d (\bi{r},t) = D_d \nabla^2 w_d(\bi{r},t),
\label{eq:appendix3}
\end{equation}
with the initial condition  
\begin{equation}
w_d (\bi{r},t=0)=1, 
\label{eq:appendix4}
\end{equation}
and the boundary conditions 
\begin{equation}
w_d(r=\sigma,t)=0,~~~~~
\lim_{r \rightarrow \infty} w_d (r,t)=1. 
\label{eq:appendix5}
\end{equation}

Since $1 - w_d (r,t)$ is the probability that the pair coalesce at time $t$, 
the mean coalescence time $\tau(r)$ is given by
\begin{equation}
\tau(r) = \int_0^\infty {\rm d}t \, t \frac{\partial}{\partial t} 
\left[ 1 - w_d (r,t) \right] 
=  \int_0^\infty {\rm d}t \, w_d (r,t). 
\label{eq:appendix6}
\end{equation}
In general, 
the mean coalescence time is called the mean first-passage time. 
Integrating Eq.~(\ref{eq:appendix3}) over time and using Eq.~(\ref{eq:appendix4}), 
we find that the mean coalescence time satisfies Eq.~(\ref{eq:mft1}) when 
domains are confined in the spherical surface. 
Note that in Eq.~(\ref{eq:appendix3}), the diffusion equation 
in the spherical surface can be written by assuming azimuthal symmetry such that 
\begin{eqnarray}
D_2 \nabla^2 w_2&=D_2 \frac{1}{R^2 \sin \theta} \frac{\partial}{\partial \theta} 
\left(
\sin \theta \frac{\partial w_2}{\partial \theta}
\right) \nonumber \\
&= \frac{D_2}{R^2} \frac{\partial}{\partial z} (1-z^2) \frac{\partial w_2}{\partial z}, 
\end{eqnarray}
where  $z=\cos \theta$.


\section*{References}
\begin{thebibliography}{10}

\bibitem{saffman-75}
Saffman P G and Delbr{\"u}ck M 1975 
Brownian motion in biological membranes
{\it Proc. Natl. Acad. Sci. USA} {\bf 72} 3111

\bibitem{saffman-76}
Saffman P G 1976 
Brownian motion in thin sheets of viscous fluid
{\it J. Fluid Mech.} {\bf 73} 593 

\bibitem{DeKoker}
R.~De Koker, {\it Domain structures and hydrodynamics in lipid monolayers. PhD dissertation},
Stanford University (1996)

\bibitem{Komura2012} 
Komura S, Ramachandran S, Seki K and Imai M 2012  
{\it Dynamics of heterogeneity in fluid membranes} 
chapter in the book {\it Advances in Planar Lipid bilayers and Liposomes} edited by Igli\v{c} A 
Vol. 16 (Academic Press, Elsevier: Amsterdam)

\bibitem{Fujitani}
Fujitani Y 2011 
Drag Coefficient of a Liquid Domain in a Fluid Membrane
{\it J. Phys. Soc. Jpn}  {\bf 80} 074609

\bibitem{Seki}
Seki K, Ramachandran S, and Komura S 2011 
Diffusion coefficient of an inclusion in a liquid membrane supported by
a solvent of arbitrary thickness 
{\it Phys. Rev. E} {\bf 84} 021905 

\bibitem{Ramachandran-10}
Ramachandran S, Komura S and Gompper G 2010 
Effects of an embedding bulk fluid on phase separation dynamics in a thin liquid film
{\it EPL} {\bf 89} 56001

\bibitem{Camley}
Camley B A and Brown F L H 2011 
Dynamic scaling in phase separation kinetics for quasi-two-dimensional membranes
{\it J. Chem. Phys.} {\bf 135} 225106 

\bibitem{Fan}
Fan J, Han T and Haataja M 2010 
Hydrodynamic effects on spinodal decomposition kinetics in planar lipid bilayer membranes 
{\it J. Chem. Phys.} {\bf 133} 235101

\bibitem{Petrov}
Petrov E P, Petrosyan R, and Schwille P 2012 
Translational and rotational diffusion of micrometer-sized solid domains in
lipid membranes  
{\it Soft Matter} {\bf 8} 7552 

\bibitem{veatch-03}
Veatch S L and Keller S L 2003
Separation of liquid phases in giant vesicles of ternary mixtures of phospholipids 
and cholesterol 
{\it Biophys. J.} {\bf 85} 3074

\bibitem{Baumgart-03}
Baumgart T, Hess S T, and Webb W. W.  2003 
Imaging coexisting fluid domains in biomembrane models coupling curvature and line tension
{\it Nature}  {\bf 425} 821

\bibitem{saeki-06}
Saeki D, Hamada T and Yoshikawa K 2006
Domain-growth kinetics in a cell-sized liposome 
{\it J. Phys. Soc. Jpn.} {\bf 75} 013602

\bibitem{yanagisawa-07}
Yanagisawa M, Imai M, Masui T, Komura S and Ohta T 2007 
Growth dynamics of domains in ternary fluid vesicles
{\it Biophys. J.} {\bf 92} 115

\bibitem{Sezgin}
Sezgin E, Kaiser HJ, Baumgart T, Schwille P, Simons K, Levental I 2012 
Elucidating membrane structure and protein behavior using giant plasma membrane vesicles 
{\it Nature Protocols} {\bf 7}  1042 

\bibitem{Bray}
Bray A J 2002 
Coarsening dynamics of phase-separating systems
{\it Adv. Phys.} {\bf 51} 481 

\bibitem{Lipowsky1992}
Lipowsky R 1992  
Budding of membranes induced by intramembrane domains 
{\it J. Phys. France II} {\bf 2} 1825

\bibitem{Lipowsky2003}
Lipowsky R and Dimova R 2003 
Domains in membranes and vesicles  
{\it J. Phys.: Condens. Matter} {\bf 15} S31

\bibitem{Seifert1997}
Seifert U 1997   
Configurations of fluid membranes and vesicles 
{\it Adv. Phys. } {\bf 46} 1825

\bibitem{Ursell2009}
Ursell T S, Klug W S and Phillips R 2009  
Morphology and interaction between lipid domains 
{\it Proc. Natl. Acad. Sci. U.S.A.} {\bf 106} 13301 

\bibitem{Semrau2009}
Semrau S, Idema T, Schmidt T and Storm C 2009 
Membrane-mediated interactions measured using membrane domains 
{\it Biophys. J.} {\bf 96} 4906 

\bibitem{Andelman}
Andelman  D, Kawakatsu T and Kawasaki K 1992 
Equilibrium shape of two-component unilamellar membranes and vesicles 
{\it Europhys. Lett.} {\bf 19}  57

\bibitem{Parthasarathy}
Parthasarathy R, Yu C H and Groves J T 2006 
Curvature-modulated phase separation in lipid bilayer membranes 
{\it Langmuir} {\bf 22} 5095 

\bibitem{Garcia-Saez}
Garc\'{i}a-S\'{a}ez A J, Chiantia S and Schwille P 2007 
Effect of line tension on the lateral organization of lipid membranes 
{\it J. Biol. Chem.} {\bf 282} 33537

\bibitem{binder-74}
Binder K and Stauffer D 1974
Theory for the slowing down of the relaxation and spinodal decomposition of 
binary mixtures
{\it Phys. Rev. Lett.} {\bf 33} 1006 

\bibitem{Tomita}
Tomita H 1976 
Statistical properties of random interface systems
{\it Prog. Theor. Phys.} {\bf 56} 1661 

\bibitem{Naqvi}
Razi Naqvi K 1974
Diffusion-controlled reactions in two-dimensional fluids: discussion of 
measurements of lateral diffusion of lipids in biological membranes
{\it Chem. Phys. Lett.} {\bf 28} 280

\bibitem{Barzykin}
Barzykin  A V,  Seki K and Tachiya M 2001 
Kinetics of diffusion-assisted reactions in microheterogeneous systems  
{\it Adv. Coll. Inter. Sci.} {\bf 89-90} 47  

\bibitem{LK04}
Laradji M and Kumar P B S 2001 
Dynamics of domain growth in self-assembled fluid vesicles
{\it Phys. Rev. Lett.} {\bf 93} 198105 

\bibitem{Sano}
Sano H and Tachiya M 1981
Theory of diffusion-controlled reactions on spherical surfaces and its 
application to reactions on micellar surfaces 
{\it J. Chem. Phys.} {\bf 75} 2870

\bibitem{Bloomfield}
Bloomfield V A and Prager S 1979 
Diffusion-controlled reactions on spherical surfaces. Application to 
bacteriophage tail fiber attachment
{\it Biophys. J.} {\bf 27} 447

\bibitem{Smoluchowski}
Smoluchowski M V 1916 
Drei Vortr\"{a}ge \"{u}ber Diffusion, Brownsche Bewegung und Koagulation von 
Kolloidteilchen
{\it Physik. Zeit.} {\bf 17} 557

\bibitem{Rice}
Rice S A 1985  
{\it Diffusion-Limited Reactions} in
{\it Comprehensive Chemical Kinetics} edited by
Bamford C H, Tipper C F H and Compton R G  
Vol.~25 (Elsevier:Amsterdam)
and references cited therein.

\bibitem{Tachiya83}
Tachiya M 1983
Theory of diffusion-controlled reactions: Formulation of the bulk reaction rate in terms of the pair probability
{\it Radiat. Phys. Chem.} {\bf 21} 167 

\bibitem{Oshanin}
Oshanin G, Moreau M and Burlatsky S F 1994 
Models of chemical reactions with participation of polymers 
{\it Adv. Colloid Interface Sci.} {\bf 49} 1  

\bibitem{Henle}
Henle M L and Levine A J 2010
Hydrodynamics in curved membranes: The effect of geometry on particulate mobility
{\it Phys. Rev. E} {\bf 81} 011905 

\bibitem{abram-stegun}
Abramowitz M and Stegun I A 1972 
{\it Handbook of Mathematical Functions} (Dover: New York)

\bibitem{Adam} 
Adam G and Delbr{\"u}ck M 1968 
{\it  Structural Chemistry and
Molecular Biology} eds. Rich A and Davidson N 
(Freeman, San Francisco) 198

\bibitem{Taniguchi}
Taniguchi T, Yanagisawa M and Imai M 2011 
Numerical investigations of the dynamics of two-component vesicles
{\it J. Phys.: Condens. Matter} {\bf 23} 284103 

\bibitem{Lifshitz}
Lifshitz E M and Pitaevskii L P 1981 
{\it Physical Kinetics} (Pergamon Press, Oxford).

\bibitem{Miguel}
Miguel M S ,Grant M, Gunton J D 1985 
Phase separation in two-dimensional binary fluids 
{\it Phys. Rev. A} {\bf 31} 1001

\bibitem{Ehrig}
Ehrig J, Petrov E P  and Schwille P 2011 
{\it New J.  Phys.}  {\bf 13} 045019 

\bibitem{Weiss}
Weiss G H 1967 
First passage time problems in chemical physics 
{\it Adv. Chem. Phys.} {\bf 13} 1.

\bibitem{Risken}
Risken H 1984 {\it The Fokker-Planck Equation: Methods of Solution and Applications} (Springer:New York). 

\bibitem{Gardiner}
Gardiner C W 1983 {\it Handbook of Stochastic Methods} (Springer:Berlin). 

\end{thebibliography}
\end{document}